\documentclass{emulateapj}

\usepackage{amsmath}

\newcommand{\about}{$\sim\!\!$~}

\newcommand{\be}{\begin{displaymath}}
\newcommand{\ee}{\end{displaymath}}

\def\lsim{\hbox{\rlap{\raise 0.425ex\hbox{$<$}}\lower 0.65ex\hbox{$\sim$}}}
\def\gsim{\hbox{\rlap{\raise 0.425ex\hbox{$>$}}\lower 0.65ex\hbox{$\sim$}}}
\def\arcmin{\hbox{$^\prime$}}

\newcommand{\kms}{km\,s$^{-1}$}

\shorttitle{The Normal Young SN~Ia 2017cfd}
\shortauthors{Han et al.}

\begin{document}

\title{SN 2017cfd: A Normal Type Ia Supernova Discovered Very Young}

\author{Xuhui Han\altaffilmark{1},          
WeiKang Zheng\altaffilmark{2},              
Benjamin E. Stahl\altaffilmark{2,3,4},      
Jamison Burke\altaffilmark{5,6},            
Jozsef Vinko\altaffilmark{7,8,9},           
Thomas de Jaeger\altaffilmark{2,10},        
Thomas G. Brink\altaffilmark{2},            
Borbala Cseh\altaffilmark{7},               
Daichi Hiramatsu\altaffilmark{5,6},         
D. Andrew Howell\altaffilmark{5,6},         
Bernadett Ignacz\altaffilmark{7},           
Reka Konyves-Toth\altaffilmark{7},          
Mate Krezinger\altaffilmark{7},             
Curtis McCully\altaffilmark{5,6},           
Andras Ordasi\altaffilmark{7},              
Dora Pinter\altaffilmark{7},                
Krisztian Sarneczky\altaffilmark{7},        
Robert Szakats\altaffilmark{7},             
Kevin Tang\altaffilmark{2},                 
Krisztian Vida\altaffilmark{7},             
Jing Wang\altaffilmark{1,11},               
Jianyan Wei\altaffilmark{1},                
J. Craig Wheeler\altaffilmark{9,12},        
Liping Xin\altaffilmark{1},                 
and Alexei V. Filippenko\altaffilmark{2,13}     
}

\altaffiltext{1}{Key Laboratory of Space Astronomy and Technology, National Astronomical Observatories, Chinese Academy of Sciences, Beijing 100101, China (email: hxh@nao.cas.cn).}
\altaffiltext{2}{Department of Astronomy, University of California, Berkeley, CA 94720-3411, USA (email: weikang@berkeley.edu).}
\altaffiltext{3}{Department of Physics, University of California, Berkeley, CA 94720, USA.}
\altaffiltext{4}{Marc J. Staley Graduate Fellow.}
\altaffiltext{5}{Las Cumbres Observatory Global Telescope Network, 6740 Cortona Dr Ste 102, Goleta, CA 93117-5575, USA.}
\altaffiltext{6}{Department of Physics, University of California, Santa Barbara, CA 93106-9530, USA.}
\altaffiltext{7}{Konkoly Observatory, Institute for Astronomy and Earth Sciences, Konkoly Thege ut 15-17, Budapest, 1121 Hungary.}
\altaffiltext{8}{Department of Optics and Quantum Electronics, University of Szeged, Dom ter 9, Szeged, 6720, Hungary.}
\altaffiltext{9}{Department of Astronomy, University of Texas, 2515 Speedway, Austin, TX 78712, USA.}
\altaffiltext{10}{Bengier Postdoctoral Fellow.}
\altaffiltext{11}{Guangxi Key Laboratory for Relativistic Astrophysics, School of Physical Science and Technology, Guangxi University, Nanning 530004, China.}
\altaffiltext{12}{George P. and Cynthia Woods Mitchell Institute for Fundamental Physics \& Astronomy, Texas A\&M University, Department of Physics and Astronomy, 4242 TAMU, College Station, TX 77843, USA.}
\altaffiltext{13}{Miller Senior Fellow, Miller Institute for Basic Research in Science, University of California, Berkeley, CA 94720, USA.}

\begin{abstract}
The Type~Ia supernova (SN~Ia) 2017cfd in IC~0511 (redshift $z = 0.01209 \pm 0.00016$)
was discovered by the Lick Observatory Supernova Search $1.6 \pm 0.7$\,d after
the fitted first-light time (FFLT; 15.2\,d before $B$-band maximum brightness).
Photometric and spectroscopic follow-up observations show that SN~2017cfd
is a typical, normal SN~Ia with a peak luminosity $M_B \approx -19.2 \pm 0.2$ mag,
$\Delta m_{15}(B) = 1.16$ mag,
and reached a $B$-band maximum \about16.8~d after the FFLT.
We estimate there to be moderately strong host-galaxy extinction
($A_V = 0.39 \pm 0.03$ mag) based on MLCS2k2 fitting.
The spectrum reveals a Si~II $\lambda$6355 velocity of $\sim11,200$\,\kms\ at peak brightness.
SN~2017cfd was discovered very young, with multiband data taken starting 2~d after the FFLT,
making it a valuable complement to the currently small sample (fewer than a dozen) 
of SNe~Ia with color data at such early times. We find that its intrinsic early-time 
$(B - V)_0$ color evolution belongs to the "blue'' population rather than to the 
distinct "red'' population.
Using the photometry, we constrain the companion star radius to be
$\lesssim2.5\,{\rm R}_\sun$, thus ruling out a red-giant companion.

\end{abstract}



\section{Introduction}\label{s:intro}

Type~Ia supernovae (SNe~Ia; see Filippenko 1997 for a review of supernova classification)
are the thermonuclear runaway explosions of carbon/oxygen white dwarfs
(see, e.g., Hillebrandt \& Niemeyer 2000 for a review).
One of the most important applications of SN~Ia is that they can be used as standardizable
candles for measurements of the expansion rate of the Universe
(Riess et al. 1998; Perlmutter et al. 1999). 
The have also provided the main initial evidence for "H$_0$ tension'' --- the discrepancy
in the values of H$_0$ measured locally and inferred from Planck (e.g., Riess et al. 2018, 2019).

There are two general favored progenitor systems for SNe~Ia:
the single-degenerate scenario (Hoyle \& Fowler 1960; Hachisu et al. 1996;
Meng et al. 2009; R\"{o}pke et al. 2012), which consists of a single white dwarf
accreting material from a companion, and the double-degenerate
scenario involving the merger of two white dwarfs (Webbink 1984;
Iben \& Tutukov 1984; Pakmor et al. 2012; R\"{o}pke et al. 2012).
However, our understanding of the progenitor systems and explosion mechanisms
remains substantially incomplete both theoretically and observationally
(see a recent review by Jha et al. 2019).

Extremely early discovery and follow-up observations
are essential for understanding the physical properties of SNe~Ia and for
revealing their progenitor systems.
For example, early-time light curves can be used to constrain the companion-star radius,
as in the cases of SN~2011fe (Nugent et al. 2011; Bloom et al. 2012), 
SN~2012cg (Silverman et al. 2012b),
SN~2013dy (Zheng et al. 2013), SN~2013gy (Holmbo et al. 2019),
SN~2014J (Goobar et al. 2014), iPTF14atg (Cao et al. 2015),
SN~2015F (Im et al. 2015),
SN~2017cbv (Hosseinzadeh et al. 2017), SN~2018oh (Li et al. 2019), and
SN~2019ein (Kawabata et al. 2019), although there are other alternatives
to explain the data (e.g., Piro \& Nakar 2013; Maeda et al. 2018; Magee et al., 2018;
Stritzinger et al. 2018a; Polin et al. 2019).
They can also be used to explore the "dark phase'' of SN~Ia, which can
last for a few hours to days between the moment of explosion and the first
observed light (Rabinak, Livne, \& Waxman 2012; Piro \& Nakar 2013, 2014),
as with SN~2014J (Goobar et al. 2014), SN~2015F (Im et al. 2015), and iPTF14pdk (Cao et al. 2016).
Optical spectra obtained shortly after explosion can be used to examine the possible unburned
material from the progenitor white dwarf, such as the \ion{C}{2} feature,
which was found significantly only in the very early spectra of SN~2013dy (Zheng et al. 2013)
and SN~2017cbv (Hosseinzadeh et al. 2017).

Observationally, numerous efforts have been conducted to discover young SNe~Ia,
with progressively more surveys joining in the past few years
(e.g., All-Sky Automated Survey for Supernovae, Asteroid Terrestrial-impact Last Alert System,
Palomar Transient Factory, Intermediate Palomar Transient Factory).
In 2011, to discover very young SNe~Ia (hours to days after explosion),
our Lick Observatory Supernova Search
(LOSS; Filippenko et al. 2001; Filippenko 2005; Leaman et al. 2011) modified
its search strategy. Since that year, using the 0.76\,m Katzman
Automatic Imaging Telescope (KAIT), our group monitored fewer
galaxies but with a higher cadence. 
Consequently, our group has discovered many young SNe in the past
few years (see Section 3.4). Among these discoveries is SN~2017cfd,
which was discovered
merely $1.6 \pm 0.5$\,d after the fitted first-light time (FFLT), and
KAIT automatically started multiband follow-up observations only minutes after discovery.
Here we present our optical photometry and spectroscopy together with an analysis.


\section{Discovery and Observations}\label{s:discovery}

SN~2017cfd was discovered in an 18~s unfiltered KAIT image
taken at 06:32:37 on 2017~March~16 (UT dates are used
throughout this paper), at $\sim19$\,mag in the $Clear$ band (close to the $R$ band; see Li et al. 2003).
It was reported to the Transient Name Server (TNS)
shortly after discovery by Halle, Zheng, \& 
Filippenko\footnote{https://wis-tns.weizmann.ac.il//object/2017cfd}.
We measure its J2000 coordinates to be 
$\alpha = 08^{\mathrm{h}}40^{\mathrm{m}}49.09^{\mathrm{s}}$, 
$\delta = +73^{\circ}29\arcmin15\farcs1$, 
with an uncertainty of $\pm 0\farcs5$ in each coordinate.
SN~2017cfd is $5\farcs8$ west and $3\farcs1$ north of
the nucleus of the host galaxy IC~0511, which has
redshift $z = 0.01209 \pm 0.00016$ (Falco et al. 1999)
and a spiral morphology (S0/a; de Vaucouleurs et al. 1991).

\begin{figure}
\centering
\includegraphics[width=.49\textwidth]{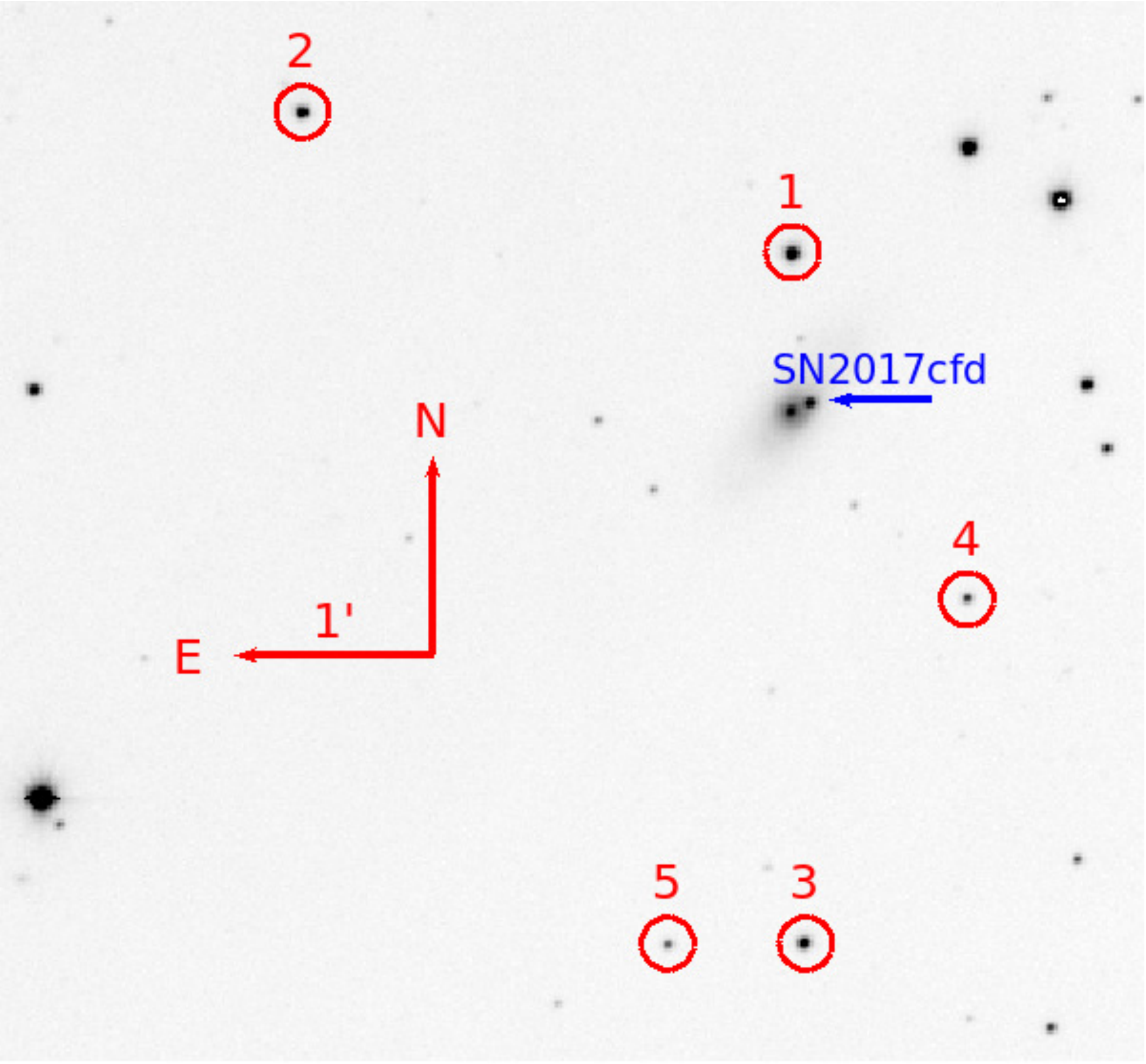}
\caption{KAIT unfiltered image showing the location of SN~2017cfd. 
Five reference stars are also marked with circles.}
\label{fig1}
\end{figure}

KAIT automatically performed multiband photometric follow-up observations of SN~2017cfd
once it was discovered. KAIT data were reduced using 
our image-reduction pipeline (Ganeshalingam et al. 2010; Stahl et al. 2019a).
We applied an image-subtraction procedure to remove the host-galaxy light,
and point-spread-function (PSF) photometry was then obtained using DAOPHOT 
(Stetson 1987) from the IDL Astronomy User's 
Library\footnote{http://idlastro.gsfc.nasa.gov/}.
The multiband data
are calibrated to local Pan-STARRS1\footnote{http://archive.stsci.edu/panstarrs/search.php}
stars (see Figure \ref{fig1}),
whose magnitudes were first transformed into the Landolt system
using the empirical prescription presented by Tonry et al. (2012, Equation 6)
and then transformed to the KAIT natural system.
Apparent magnitudes were all measured in the KAIT4 natural system.
The final results were transformed to the standard system using local
calibrators and color terms for KAIT4 (Stahl et al. 2019a).

Note that KAIT photometry of SN~2017cfd has already been published by
Stahl et al. (2019a). The major difference between our reanalysis presented here compared 
to the results of Stahl et al. (2019a) is that we preformed a more careful analysis of the
data obtained on the discovery night, where multiple short exposures
were taken for each filter. Considering that the SN was very faint ($B\approx 19.2$ mag) on
the discovery night, here we coadd the short-exposure images in each filter
in order to increase the signal-to-noise ratio before applying subtraction
and further analysis, whereas the Stahl et al.
pipeline\footnote{https://github.com/benstahl92/LOSSPhotPypeline}
did the photometry on each single image and then took the average.
We also added two epochs of upper-limit measurements in the $Clear$ band prior to discovery.

Additional photometric data were obtained with the 0.6/0.9\,m Schmidt telescope, equipped with a front-illuminated FLI Proline PL16801 
$4096 \times 4096$ pixel CCD and Johnson-Cousins $BVRI$ filters, at Piszk\'estet{\H o} Mountain Station of Konkoly Observatory, between 
2017-03-17 and 2017-05-28. The CCD frames were reduced using standard 
{\it IRAF}\footnote{IRAF is distributed by the National Optical Astronomy
Observatories, which are operated by the Association of Universities for Research in Astronomy, 
Inc., under cooperative agreement with the US National Science Foundation.}
tasks.
Template frames were taken with the same instrument and setup on 2019-03-23 and 2019-03-24, 2\,yr after peak brightness, when the SN had faded sufficiently 
below the detection limit. Subtraction of the templates was computed using self-developed {\it IRAF} scripts. After that, PSF photometry
of the SN was performed on the subtracted frames, while the determination of the PSF on each frame and photometry of local comparison stars were 
done on the original (dark- and flatfield-corrected) frames. Finally, the instrumental magnitudes were transformed to the standard Johnson-Cousins
system using PS1-photometric data\footnote{http://archive.stsci.edu/panstarrs/search.php} for the local comparison stars.

Photometric reduction for the Las Cumbres Observatory (LCO; Brown et al. 2013)
images was accomplished using
lcogtsnpipe (Valenti et al. 2016), a PyRAF-based pipeline.
Image subtraction was accomplished using PyZOGY (Guevel \& Hosseinzadeh 2017),
an implementation in Python of the subtraction algorithm described
by Zackay et al. (2016).

\begin{deluxetable}{lcccccccccc}
 \tabcolsep 0.4mm
 \tablewidth{0pt}
 \tabletypesize{\scriptsize}
 \tablecaption{Multiband Photometry of SN~2017cfd
               \label{tab:prompt_spec_par}}
 \tablehead{ \colhead{MJD} & \colhead{Mag} & \colhead{$1\sigma$} & \colhead{Mag} & \colhead{$1\sigma$} & \colhead{Mag} & \colhead{$1\sigma$} & \colhead{Mag} & \colhead{$1\sigma$} & \colhead{Mag} & \colhead{$1\sigma$} }
 \startdata
From KAIT       &\multicolumn{2}{c}{$B$}&\multicolumn{2}{c}{$V$}&\multicolumn{2}{c}{$R$}&\multicolumn{2}{c}{$I$}&\multicolumn{2}{c}{$Clear$}\\
\hline                                                                                          
\\                                                                                              
57825.322       & -     & -    & -     & -    & -     & -    & -     & -     & $>$19.0 & -      \\
57826.310       & -     & -    & -     & -    & -     & -    & -     & -     & $>$19.1 & -      \\
57828.278       & 19.24 & 0.36 & 19.09 & 0.17 & -     & -    & -     & -     & 19.10   & 0.26   \\
57829.172       & 17.71 & 0.14 & 18.69 & 0.76 & 16.87 & 0.15 & 18.37 & 0.27  & 17.82   & 0.13   \\
57830.322       & 17.06 & 0.05 & 17.09 & 0.06 & 16.80 & 0.07 & 16.38 & 0.10  & 17.16   & 0.10   \\
57831.279       & 17.18 & 0.24 & 16.91 & 0.22 & 16.29 & 0.18 & 16.68 & 0.39  & 16.57   & 0.09   \\
57841.275       & 14.81 & 0.19 & 14.97 & 0.02 & -     & -    & -     & -     & 14.68   & 0.02   \\
57843.241       & -     & -    & -     & -    & -     & -    & -     & -     & 14.67   & 0.02   \\
57845.232       & 14.96 & 0.01 & 14.84 & 0.01 & 14.73 & 0.03 & 15.05 & 0.02  & 14.82   & 0.03   \\
57864.221       & -     & -    & -     & -    & -     & -    & -     & -     & 15.88   & 0.06   \\
57872.189       & -     & -    & -     & -    & -     & -    & -     & -     & 16.31   & 0.10   \\
57840.311       & 15.06 & 0.01 & 14.98 & 0.01 & 14.77 & 0.02 & 14.85 & 0.02  & 15.07   & 0.06   \\
57844.297       & 14.98 & 0.01 & 14.86 & 0.01 & 14.72 & 0.01 & 14.99 & 0.03  & 14.68   & 0.02   \\
57849.259       & -     & -    & -     & -    & -     & -    & -     & -     & 15.25   & 0.14   \\
57853.276       & 15.57 & 0.22 & 15.16 & 0.02 & 15.19 & 0.03 & 15.32 & 0.06  & 15.44   & 0.06   \\
57878.204       & -     & -    & -     & -    & -     & -    & -     & -     & 16.73   & 0.16   \\
57881.215       & 18.82 & 0.22 & 16.85 & 0.08 & 15.92 & 0.06 & 15.56 & 0.06  & 17.04   & 0.18   \\
57883.207       & 18.52 & 0.17 & 17.04 & 0.07 & 16.28 & 0.24 & 15.58 & 0.07  & 16.96   & 0.13   \\
57887.185       & 19.02 & 0.28 & 17.36 & 0.23 & -     & -    & -     & -     & -       & -      \\
57889.197       & 19.35 & 0.24 & 17.29 & 0.12 & 16.65 & 0.11 & 15.83 & 0.14  & -       & -      \\
57891.200       & 18.80 & 0.14 & 17.68 & 0.11 & 16.62 & 0.06 & -     & -     & 17.30   & 0.12   \\
57893.194       & -     & -    & -     & -    & -     & -    & -     & -     & 17.30   & 0.18   \\
\\
\hline          
From Konkoly    &\multicolumn{2}{c}{$B$}&\multicolumn{2}{c}{$V$}&\multicolumn{2}{c}{$R$}&\multicolumn{2}{c}{$I$}&\multicolumn{2}{c}{}\\
\hline                                                                                          
\\                                                                                              
57829.020       & 17.96 & 0.07 & 17.92 & 0.06 & 17.79 & 0.05 & 17.41   & 0.07   &       &        \\
57832.880       & 16.16 & 0.02 & 16.17 & 0.02 & 15.90 & 0.02 & 15.87   & 0.03   &       &        \\
57833.900       & 15.87 & 0.03 & 15.85 & 0.02 & 15.60 & 0.02 & 15.58   & 0.03   &       &        \\
57835.860       & 15.42 & 0.02 & 15.43 & 0.02 & 15.23 & 0.03 & 15.15   & 0.03   &       &        \\
57837.820       & 15.19 & 0.04 & 15.19 & 0.03 & 14.95 & 0.03 & 14.93   & 0.03   &       &        \\
57841.860       & 14.91 & 0.03 & 14.86 & 0.02 & 14.71 & 0.02 & 14.90   & 0.02   &       &        \\
57849.850       & 15.20 & 0.05 & 14.94 & 0.04 & 14.79 & 0.08 & 14.97   & 0.24   &       &        \\
57852.830       & 15.46 & 0.03 & 15.13 & 0.02 & 15.10 & 0.04 & 15.61   & 0.06   &       &        \\
57853.830       & 15.51 & 0.03 & 15.22 & 0.02 & 15.17 & 0.03 & 15.65   & 0.05   &       &        \\
57857.820       & 15.95 & 0.03 & 15.42 & 0.01 & 15.41 & 0.03 & 15.62   & 0.03   &       &        \\
57859.830       & 16.16 & 0.04 & 15.54 & 0.02 & 15.55 & 0.04 & 15.58   & 0.03   &       &        \\
57860.830       & 16.29 & 0.04 & 15.60 & 0.02 & 15.51 & 0.03 & 15.56   & 0.03   &       &        \\
57867.830       & 17.11 & 0.03 & 15.96 & 0.02 & 15.70 & 0.05 & 15.49   & 0.04   &       &        \\
57873.810       & 17.61 & 0.04 & 16.33 & 0.02 & 15.87 & 0.02 & 15.52   & 0.03   &       &        \\
57875.840       & 17.79 & 0.03 & 16.48 & 0.02 & 16.01 & 0.02 & 15.67   & 0.03   &       &        \\
57877.820       & 17.92 & 0.03 & 16.64 & 0.03 & 16.16 & 0.03 & 15.86   & 0.04   &       &        \\
57883.840       & 18.02 & 0.08 & 16.93 & 0.04 & 16.48 & 0.04 & 16.16   & 0.03   &       &        \\
57889.860       & 18.25 & 0.04 & 17.09 & 0.02 & 16.71 & 0.03 & 16.41   & 0.03   &       &        \\
57899.940       & 18.52 & 0.06 & 17.25 & 0.04 & 16.97 & 0.03 & 17.10   & 0.07   &       &        \\
57901.830       & 18.32 & 0.03 & 17.49 & 0.02 & 17.19 & 0.03 & 17.03   & 0.04   &       &        \\
\\
\hline          
From LCO        &\multicolumn{2}{c}{$B$}&\multicolumn{2}{c}{$V$}&\multicolumn{2}{c}{$g$}&\multicolumn{2}{c}{$r$}&\multicolumn{2}{c}{$i$}\\
\hline          
\\
57833.301       & 16.10 & 0.01 & 16.07 & 0.01 & 16.00 & 0.01 & 16.03   & 0.02   & 16.21 & 0.01  \\
57839.238       & 15.11 & 0.02 & -     & -    & -     & -    & -       & -      & -     & -     \\
57840.193       & 15.05 & 0.02 & 14.98 & 0.01 & 14.91 & 0.01 & 15.02   & 0.02   & 15.36 & 0.02  \\
57844.262       & 14.98 & 0.02 & 14.82 & 0.02 & 14.78 & 0.01 & 14.85   & 0.01   & 15.51 & 0.02  \\
57849.256       & 15.17 & 0.01 & 14.91 & 0.01 & 14.96 & 0.02 & 15.00   & 0.02   & 15.72 & 0.03  \\
57850.155       & 15.28 & 0.01 & 14.98 & 0.01 & 15.04 & 0.01 & 15.07   & 0.02   & 15.79 & 0.02  \\
57860.225       & 16.33 & 0.03 & 15.57 & 0.03 & 15.80 & 0.01 & 15.68   & 0.01   & 16.35 & 0.01  \\
57866.179       & 17.02 & 0.02 & 15.92 & 0.02 & 16.66 & 0.02 & 15.83   & 0.01   & 16.20 & 0.02  \\
57871.193       & 17.42 & 0.02 & 16.25 & 0.02 & 17.06 & 0.02 & 15.96   & 0.02   & 16.11 & 0.04  \\
57875.186       & 17.73 & 0.01 & 16.55 & 0.01 & 17.41 & 0.01 & 16.30   & 0.02   & 16.34 & 0.02  \\
57887.124       & 18.15 & 0.05 & 17.19 & 0.01 & 17.89 & 0.02 & 17.06   & 0.02   & 17.26 & 0.01  \\
57893.117       & 18.30 & 0.09 & 17.43 & 0.03 & 18.01 & 0.02 & 17.32   & 0.01   & 17.67 & 0.02  \\
\enddata
\end{deluxetable}

\clearpage
In addition, SN~2017cfd was observed by the Foundation Supernova Survey in the
$g$, $r$, $i$, and $z$  filters for 6 epochs and published by Foley et al. (2018);
we include these data in our light-curve analysis.

We performed spectroscopic follow-up observations of SN~2017cfd,
with a total of 12 spectra obtained ranging from
3.5~d to 80~d after the FFLT ($-13.2$~d to +62.8~d relative to $B$-band maximum brightness).
The spectra were taken mainly with the Kast double spectrograph 
(Miller \& Stone 1993) on the Shane 3\,m telescope at 
Lick Observatory and the FLOYDS robotic spectrograph on the 
LCO 2.0\,m Faulkes Telescope North on Haleakala, Hawaii.
A single additional spectrum, which is also the earliest one, was taken with the
Low Resolution Spectrograph-2 (LRS2; Chonis et al. 2014, 2016) on the
10\,m Hobby-Eberly Telescope at McDonald Observatory.

Data were reduced following standard techniques for CCD processing and spectrum extraction using {\it IRAF}.
The spectra were flux calibrated through observations of appropriate spectrophotometric standard stars.
All Kast spectra were taken at or near the parallactic angle
(Filippenko 1982) to minimize differential light losses caused by
atmospheric dispersion, and were reduced using
KastShiv\footnote{https://github.com/ishivvers/TheKastShiv} pipeline. 
Low-order polynomial fits to calibration-lamp spectra were used to
determine the wavelength scale, and small adjustments derived from 
night-sky emission lines in the target frames were applied. 
Flux calibration and telluric-band removal were done with our own
IDL routines; details are described by Silverman et al. (2012a) and
Shivvers et al. (2019).

\section{Light-Curve Analysis}\label{s:lightcurves}
\begin{figure*}
\centering
\includegraphics[width=0.525\textwidth]{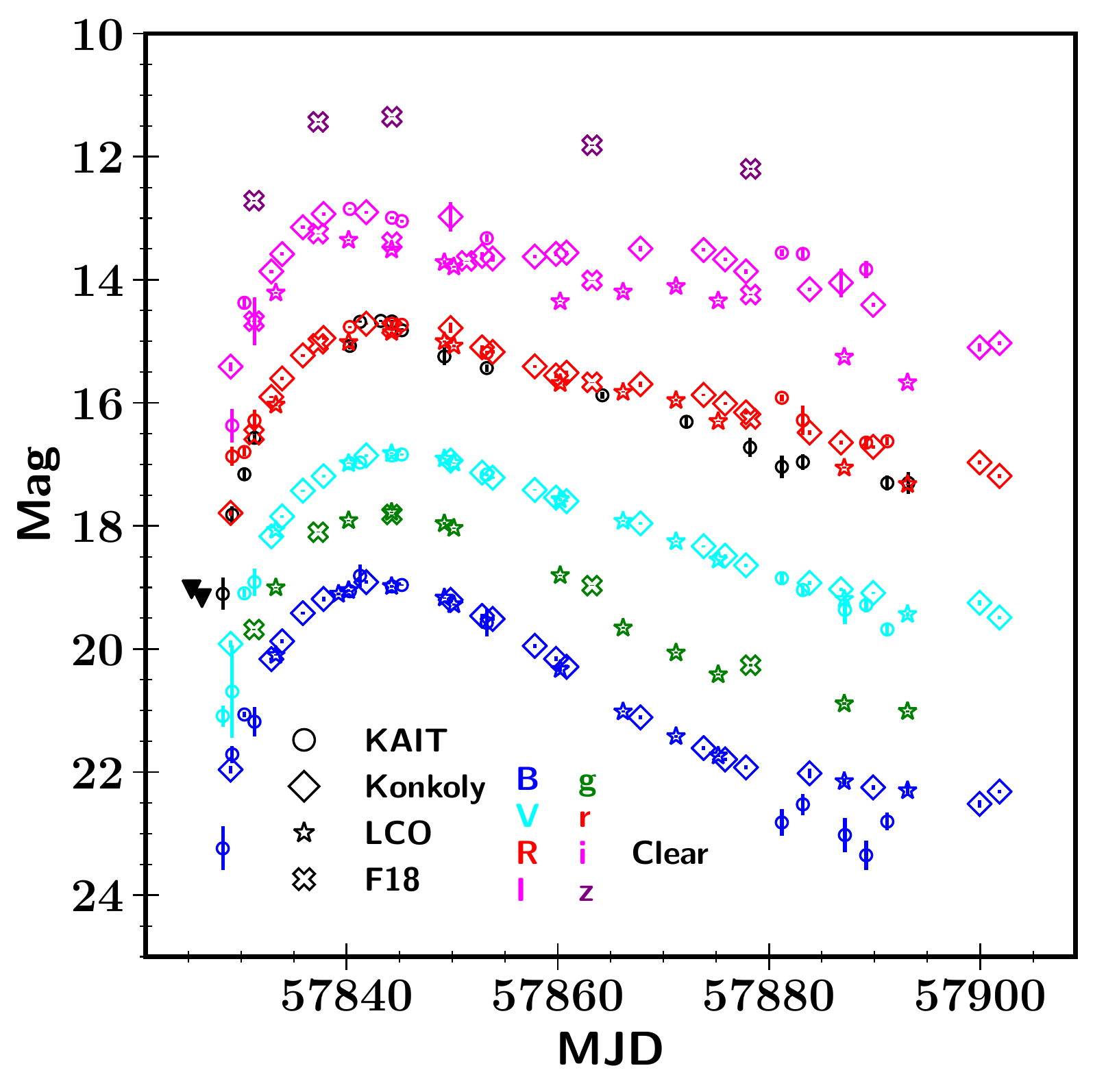}
\includegraphics[width=0.455\textwidth]{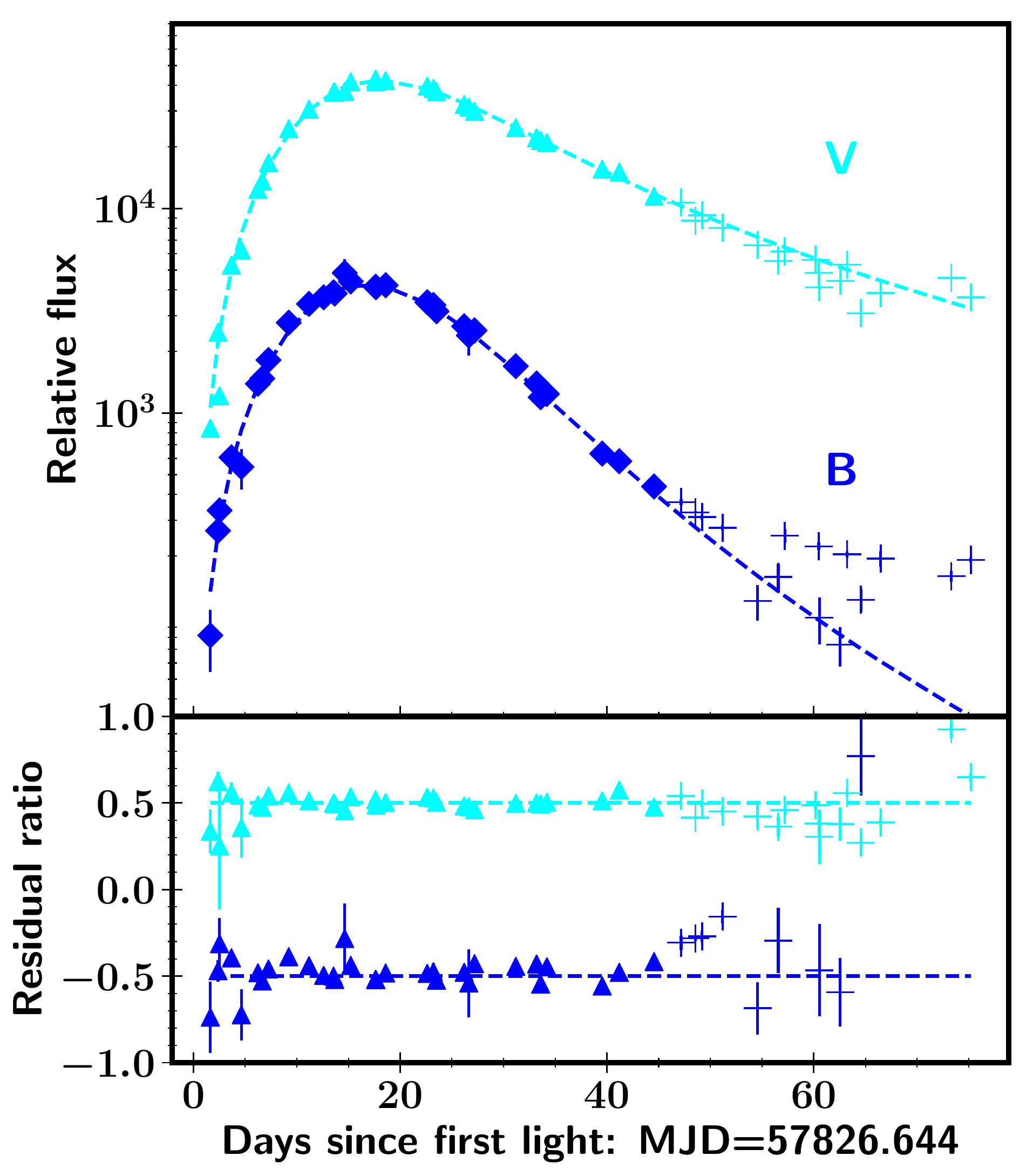}
\caption{Left: Light curve of SN~2017cfd from KAIT, LCO, Konkoly, and
         the Foundation Supernova Survey observations (labeled as F18); colors and symbols are
         coded for different filters and sources. The two black solid triangles mark the
         nondetections from KAIT $Clear$ observations one and two days before discovery.
         Right: $B$ and $V$ light-curve fitting using the analytic function presented
         by Zheng \& Filippenko (2017a, Equation 7) from discovery to $\sim 45$ days later.
         Solid data points are included in the fit while cross-shaped ones are excluded.
  }
\label{fig2}
\end{figure*}

The left panel of Figure \ref{fig2} shows the multiband light curves of SN~2017cfd
from KAIT, LCO, Konkoly, and the Foundation Supernova Survey observations;
colors and symbols are coded for different sources, with photometric data listed in Table 1.
Data in the $BVR$ and $Clear$ filters are given in the Vega system, while those in the
$gri$ filters are given in the AB system.
As one can see, we have full photometric coverage from discovery to $\sim 80$ days thereafter
in six optical bands. 
The light curves show that SN~2017cfd was discovered at a very early time, with its discovery magnitude
in the $B$ band $> 4$ mag below peak brightness.
Applying a low-order polynomial fit, we find that SN~2017cfd reached an
apparent peak of $14.95 \pm 0.03$ mag at MJD $= 57843.42$ in $B$,
and $\Delta m_{15}(B) = 1.16 \pm 0.11$ mag.

\subsection{Estimating the First-Light Time}\label{ss:first_light_time}
To determine the first-light time $t_0$ (note that here we find the
first-light time rather than the explosion time since the SN may
exhibit a "dark phase''), we use a broken-power-law function, presented
as Equation 7 of Zheng \& Filippenko (2017a), to fit the light curve
from the discovery date to $\sim 45$ days later.
Such a function was shown to be mathematically analytic
and physically related to the SN parameters (Zheng \& Filippenko 2017a).
The function has also been applied to fit $B$ light curves of 56 SNe~Ia 
(Zheng et al. 2017b) and $R$ light curves of 256 SNe~Ia of $R$-band (Papadogiannakis et al. 2019).
The light curves of SN~2017cfd were fit with all parameters free, but only the
$B$ and $V$ bands, which have the
best coverage and also to avoid the second peak in the redder bands.
The right panel of Figure \ref{fig2} shows the best-fit result; we
find the FFLT $t_{0(B)} = 57826.64 \pm 0.7$ and
$t_{0(V)} = 57826.98 \pm 0.7$, consistent with each other.
Since Zheng et al. (2017b) showed with a large sample that the FFLT
has smaller scatter in $B$ than in $V$ or other bands, here we
adopt the $B$-band result ($t_0 = 57826.64 \pm 0.7$) for later analysis.

With the FFLT and peak-time values derived above, we estimate the SN~2017cfd rise time
to be 16.8 days, very typical for SNe~Ia (Zheng et al. 2017b).
It also means that the SN was discovered merely 1.6\,d after the FFLT,
15.2\,d before $B$-band maximum light.
This makes SN~2017cfd one of the earliest detected SNe~Ia in addition to
other others, such as
SN~2009ig (Foley et al. 2012), SN~2011fe (Nugent et al. 2011),
SN~2012cg (Silverman et al. 2012b), SN~2013dy (Zheng et al. 2013),
SN~2013gy (Holmbo et al. 2019), SN~2014J (Zheng et al. 2014; Goobar et al. 2014),
iPTF14atg (Cao et al. 2015), SN~2015F (Im et al. 2015),
SN~2017cbv (Hosseinzadeh et al. 2017), SN~2018oh (Li et al. 2019), and
SN~2019ein (Kawabata et al. 2019).

\subsection{Distance and Extinction}\label{ss:distanceandextinction}

Adopting a standard cosmological model with H$_0 = 73$ km s$^{-1}$ Mpc$^{-1}$,
$\Omega_M = 0.27$, and $\Omega_{\Lambda} = 0.73$,
as well as $z = 0.01209$, a distance modulus of $33.52 \pm 0.15$ mag
(here labeled as $\mu_1$) is obtained.
With $E(B-V)_\textrm{MW} = 0.02$\,mag (Schlafly \& Finkbeiner 2011),
this implies that SN~2017cfd has an absolute magnitude of $M_B = -18.6 \pm 0.2$\,mag at peak brightness
before correcting for host-galaxy extinction. This is somewhat fainter
than the typical, normal SN~Ia (we expect $M_B\approx -19.4$\,mag from
the Phillips (1993) relation with the above value of $\Delta m_{15}(B)$);
Thus, SN~2017cfd likely suffered a certain amount of host-galaxy
extinction. In fact, from the spectra (see Section 4),
we clearly see the \ion{Na}{1}~D absorption feature, which is often converted
into reddening (but with large scatter) based on empirical relationships
(Poznanski et al. 2011; Stritzinger et al. 2018b). In Section 4, using the equivalent width (EW)
of \ion{Na}{1}~D, we estimate an extinction of $A_V = 1.34 \pm 0.40$ mag and
$E(B-V) = 0.45 \pm 0.13$ assuming $R_V = 3.1$.
This amount of host extinction appears too high for SN~2017cfd, as discussed
above; SN~2017cfd would have a peak absolute magnitude
too bright for a normal SN~Ia according to the Phillips (1993) relation.

In order to obtain an independent estimate of the host extinction, 
we performed a MLCS2k2 (Jha et al. 2007)
fit to the $B$, $V$, $R$, and $I$ light curves (not including $g$, $r$, $i$, and $z$ as there are
no template light curves in the default MLCS2k2 settings) of SN~2017cfd by fixing $R_V = 1.7$
(there are indications that $R_V = 3.1$ overestimates the host-galaxy extinction; 
e.g., Hicken et al. 2009).
The fitting parameters are given in Table 2; we found $\Delta = 0.03 \pm 0.03$,
a peak-brightness time of $57843.41 \pm 0.08$ (consistent with the value
derived from the low-order polynomial fit), and $A_V = 0.39 \pm 0.03$ mag.
With this amount of host extinction correction, SN~2017cfd has a peak absolute magnitude
of $M_B = -19.2 \pm 0.2$ mag, which is now consistent with the expectation from
the Phillips relation. 

We also performed a SALT2 (Guy et al. 2007) fitting to the
SN~2017cfd light curves. To directly compare with the MLCS2k2 fitting, we first applied
the SALT2 fitting with the $B$, $V$, $R$, and $I$ light curves. We then applied
a second SALT2 fitting to all the $BVRIgri$ light curves (as shown in
Figure 3); the SALT2 fitting parameters are given in Table 2.
As one can see in Figure 3, the SALT2 model provides very good
fits to the light curves of SN~2017cfd, and the fitting results are consistent with
those of MLCS2k2. The peak time derived from MLCS2k2 fitting and two SALT2 fittings
are within 0.3 days, and the distance moduli (all shifted to H$_0 = 73$ km s$^{-1}$ Mpc$^{-1}$)
derived from MLCS2k2 fitting ($\mu_2$) and two SALT2 fittings ($\mu_3$ and $\mu_4$) are
within the 1$\sigma$ uncertainties; they are also consistent with $\mu_1$.

\begin{figure}
\centering
\includegraphics[width=.48\textwidth]{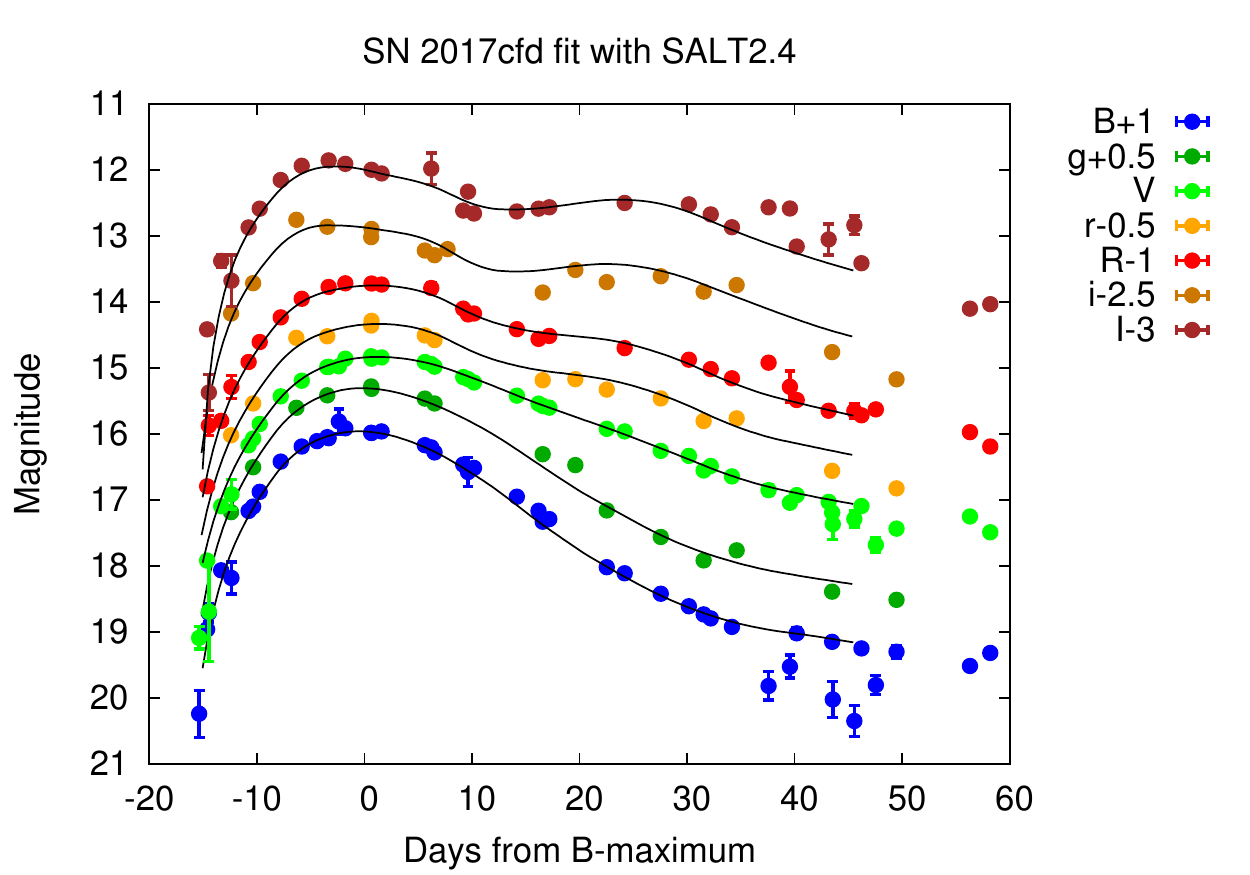}
\caption{SALT2 fitting results to the $BVRIgri$ light curves of SN~2017cfd.
  }
\label{fig3}
\end{figure}

\begin{deluxetable*}{cc|cc|cc}
 \tabcolsep 0.4mm
 \tablewidth{0pt}
 \tabletypesize{\scriptsize}
 \tablecaption{MLCS2k2 and SALT2 Fitting Results
               \label{tab:mlcs2k2_salt2_fit_result}}
 \tablehead{ \colhead{parameter} & \colhead{value} & \colhead{parameter} & \colhead{value} & \colhead{parameter} & \colhead{value} }
 \startdata
\\
\multicolumn{2}{c} {MLCS2k2 ($B,V,R,I$)} & \multicolumn{2}{c} {SALT2 ($B,V,R,I$)} &  \multicolumn{2}{c} {SALT2 ($B,V,R,I,g,r,i$)}  \\
\\
$\mu_2$ (mag)     &  33.59$\pm$0.05    &  $\mu_3$ (mag)      &  33.58$\pm$0.09     &   $\mu_4$ (mag)      &  33.70$\pm$0.07    \\
Peak time   &  57843.41$\pm$0.08 &  Peak time    &  57843.76$\pm$0.03  &   Peak time    &  54843.64$\pm$0.03 \\
$\Delta$    &  0.03$\pm$0.03     &  C            &  0.0804$\pm$0.0253  &   C            &  0.0375$\pm$0.0177 \\
$A_V$ (mag)      &  0.39$\pm$0.03     &  x0           &  0.0200$\pm$0.0005  &   x0           &  0.0203$\pm$0.0004 \\
            &                    &  x1           &  -0.6149$\pm$0.0307 &   x1           &  -0.6005$\pm$0.0240
\enddata
\end{deluxetable*}

Comparing the \ion{Na}{1}~D EW and MLCS2k2 fitting methods for estimating
the host extinction, MLCS2k2 appears to have a more reasonable
result; considering
that the \ion{Na}{1}~D EW method has large scatter, we adopt
a host extinction of $A_V = 0.39 \pm 0.03$ mag with $R_V = 1.7$ as our final result for
further analysis. Thus, SN~2017cfd has a peak absolute magnitude of
$M_B = -19.2 \pm 0.2$ mag (adopting the $\mu_1$ value),
after correcting for both the Milky Way extinction and the host extinction.

With a peak magnitude $M_B = -19.2 \pm 0.2$ mag, a rise time of 16.8\,d, and a
\ion{Si}{2} $\lambda$6355 velocity of $\sim11,200$\,\kms\ (see Section 4) at peak brightness,
we derive $M_{v^2t^2}$ (as used by Zheng et al. 2018) to be $-11.37$, thus
putting SN~2017cfd slightly under, but roughly consistent with (given the 2$\sigma$ uncertainties), 
the  $M_p$ vs. $M_{v^2t^2}$ relationship presented by Zheng et al.
(2018; see the right panel of their Figure 6).

\subsection{Pseudobolometric Light Curve}\label{ss:pseudo-bolometriclightcurve}

\begin{figure}
\centering
\includegraphics[width=.48\textwidth]{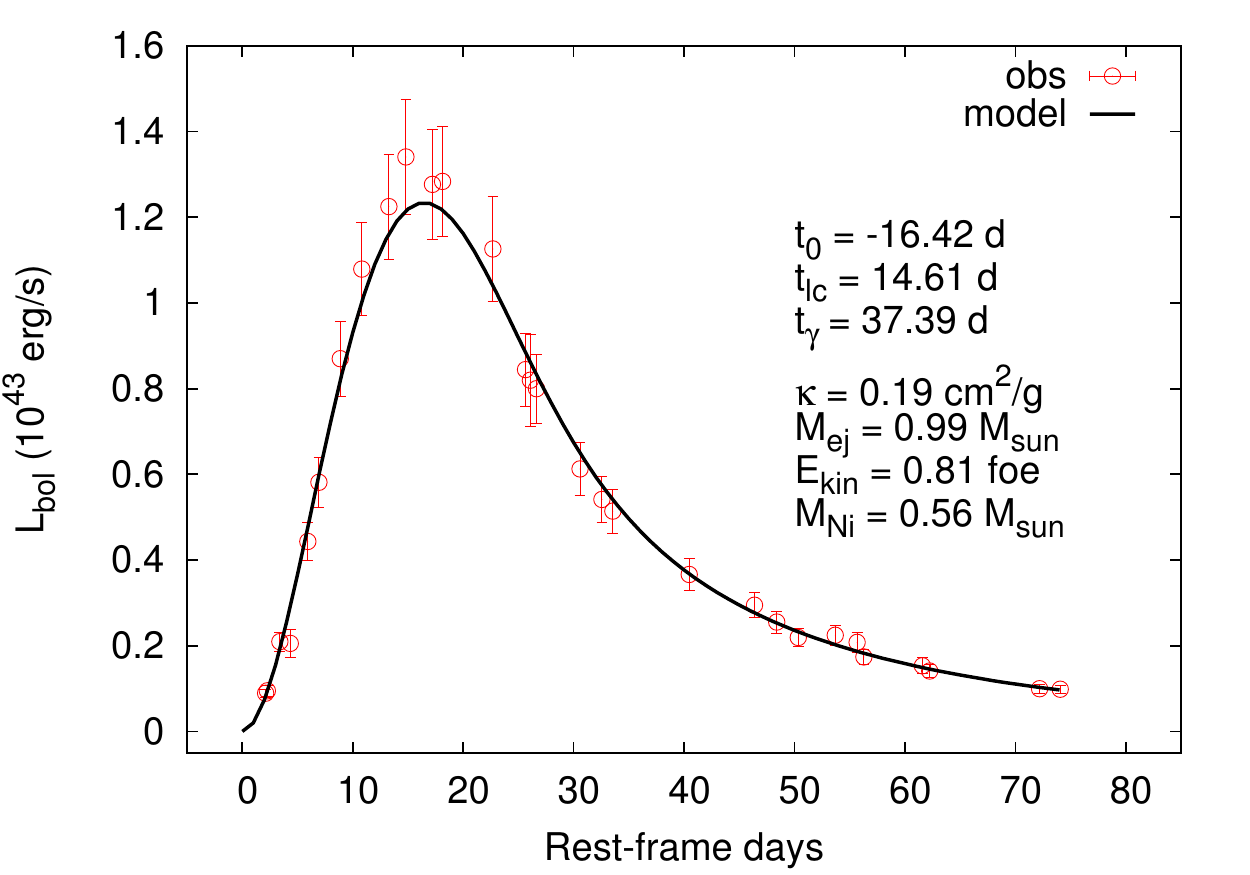}
\caption{Pseudobolometric light-curve fitting of SN~2017cfd assembled using $BVRI$
         photometric data after correcting for redshift, interstellar extinction, and distance;
         see text for details.
  }
\label{fig4}
\end{figure}

The pseudobolometric light curve of SN 2017cfd was assembled using our $BVRI$
photometric data after correcting for redshift, interstellar extinction, and distance.
We apply the same procedure as Li et al. (2019) used for SN~2018oh, except that the missing
ultraviolet fluxes are estimated by assuming zero flux at 2000\,\AA\ and a linear flux increase 
between this wavelength and the mid-wavelength of the $B$ band. The unobserved infrared fluxes are
approximated by attaching a Rayleigh-Jeans tail to the observed $I$-band flux and integrating
it between the wavelength of the $I$ band and infinity. These approximations are found to be
reasonably good representations for the unobserved ultraviolet and infrared parts of the 
spectral energy distribution (SED) of SNe~Ia (Konyves-Toth et al., in prep.).  

The pseudobolometric light curve is fit by a modified Arnett model including partial gamma-ray
leaking from the diluting ejecta, assuming $\kappa_\gamma = 0.03$ cm$^2$ g$^{-1}$
(see Li et al. 2019 for details). The best-fit model, found by $\chi^2$-minimization,
has the following parameters: rest-frame days between FFLT and $B$-band maximum
$t_r = 16.42 \pm 0.06$ days, light-curve timescale $t_d = 14.61 \pm 0.27$ days, $\gamma$-ray
leakage timescale $t_\gamma = 37.39 \pm 0.66$ days, initial mass of radioactive
$^{56}$Ni $M_{\rm Ni} = 0.56 \pm 0.05$ M$_\odot$, where the uncertainty of $M_{\rm Ni}$ also
contains the estimated uncertainty of the distance modulus ($\sim 0.1$ mag;
see Section 3.2). Following Li et al. (2019), we find
$\kappa = 0.19$ cm$^2$ g$^{-1}$, $M_{\rm ej} = 0.99$ M$_\odot$, and $E_{\rm kin} = 0.81$ for
the mean opacity, ejecta mass, and kinetic energy, respectively. These values are within
the range of typical SNe~Ia as recently found by Scalzo et al. (2019) for a larger sample.
From $M_{\rm ej}$ and $E_{\rm kin}$, the average scaling velocity is $\sim 11,700$ km~s$^{-1}$,
which agrees well with the observed \ion{Si}{2} expansion velocity around maximum light
($\sim 11,200$ km~s$^{-1}$; see Section 4). 

\subsection{Early-Time Color Evolution}\label{ss:earlytimecolor}

\begin{figure}
\centering
\includegraphics[width=.48\textwidth]{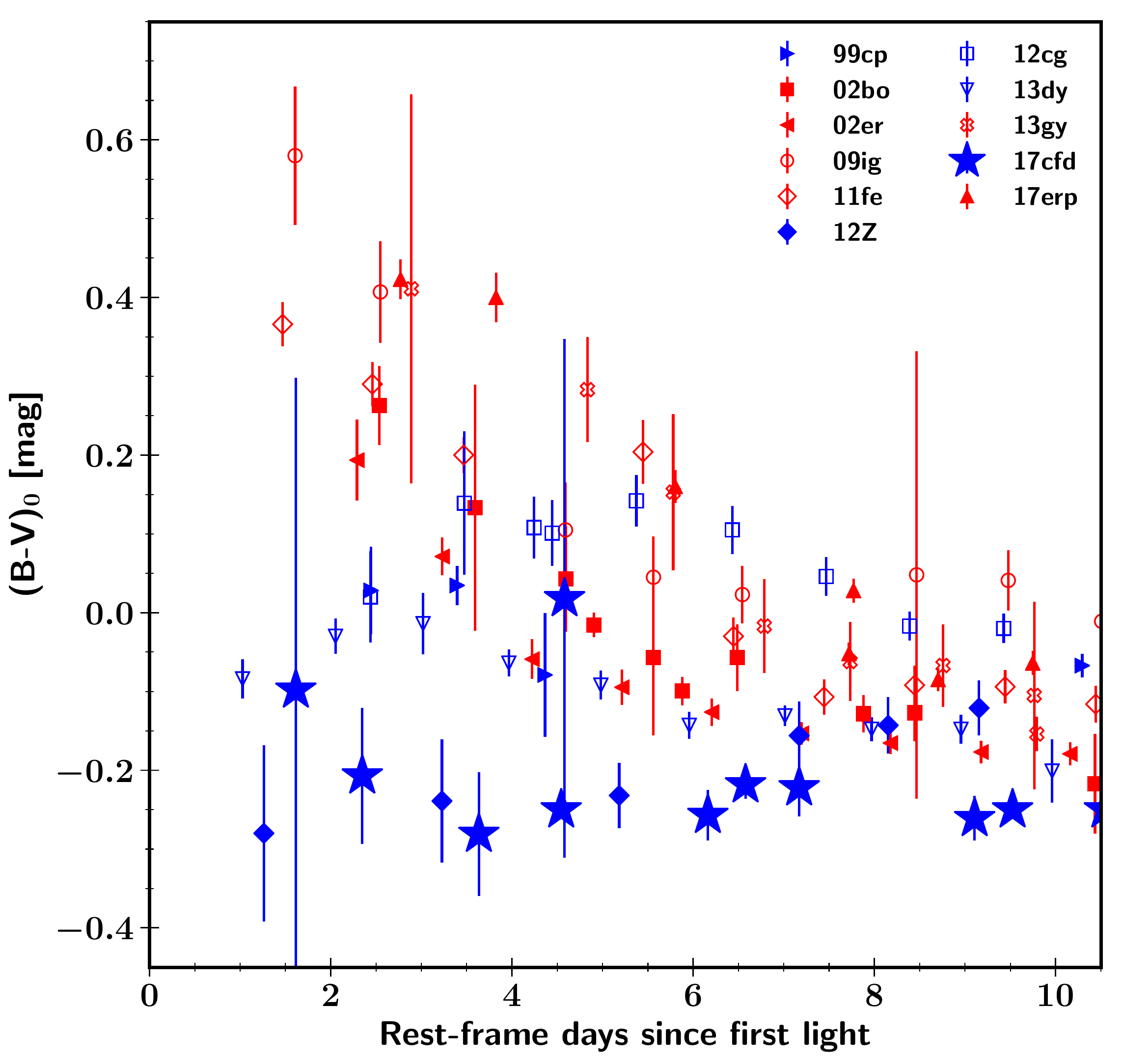}
\caption{Similar to Figure 2 of Stritzinger et al. (2018a), showing the optical $(B - V)_0$
         color evolution of SNe~Ia discovered very young. Here we replot with a subset of
         the Stritzinger et al. (2018a) sample of those SN~Ia that were discovered or
         observed by LOSS, including SN~2009ig, SN~2011fe, SN~2012cg, SN~2013dy, and SN~2013gy.
         We add six more from the LOSS sample (SN~1999cp, SN~2002bo, SN~2002er, SN~2012Z, SN~2017cfd,
         and SN~2017erp) and confirm that the
         two distinct early populations ("red'' verses "blue'') presented by Stritzinger
         et al. (2018a) remain valid. SN~2017cfd belongs to the "blue'' population.
  }
\label{fig5}
\end{figure}

\begin{deluxetable*}{lllccrcc}
 \tabcolsep 0.4mm
 \tablewidth{0pt}
 \tabletypesize{\scriptsize}
 \tablecaption{Early Color Evolution Parameters for Six SNe~Ia
               \label{tab:new_early_color_sample}}
 \tablehead{ \colhead{SN(data ref.)} & \colhead{Host} & \colhead{Redshift} & \colhead{$E(B-V)_{\rm MW}$} & \colhead{$E(B-V)_{\rm host}$} & \colhead{$t_{\rm first}$ (JD or MJD)} & \colhead{Type} & \colhead{Color} }
 \startdata
1999cp(1)   &  NGC 5468  & 0.0103  &  0.025   &  0.022$^{2}$  &  2451346.3$^{2}$   &  normal   &  blue  \\
2002bo(1,3) &  NGC 3190  & 0.0053  &  0.027   &  0.430$^{3}$  &  2452340.9$^{2}$   &  normal   &  red   \\
2002er(1)   &  UGC 10743 & 0.0090  &  0.142   &  0.218$^{4}$  &  2452508.4$^{2}$   &  normal   &  red   \\
2012Z(5)    &  NGC 1309  & 0.0071  &  0.035   &  0.070$^{6}$  &    55953.9$^{7}$   &  Iax      &  blue  \\
2017cfd(8)  &  IC  0511  & 0.01209 &  0.020   &  0.230$^{8}$  &    57826.6$^{8}$   &  normal   &  blue  \\
2017erp(5)  &  NGC 5861  & 0.00674 &  0.095   &  0.150$^{9}$  &    57916.5$^{7}$   &  normal   &  red
\enddata
\tablenotetext{}{(1) Ganeshalingam et al. (2010), (2) Zheng et al. (2017b), (3) Benetti et al. (2004), (4) Pignata et al. (2004),
                 (5) Stahl et al. (2019a), (6) Stritzinger et al. (2015), (7) This work, using the method given by Zheng et al. (2017b),
                 (8) This paper, (9) Brown et al. (2019). Reddening values are given in mag.}
\end{deluxetable*}

Following the discovery of SN~2017cfd shortly after explosion, KAIT was able to immediately 
obtain multiband data including $B$, $V$, $Clear$ (similar to $R$), and $I$, thereby
providing early-time colors.

Stritzinger et al. (2018a) found that there are two distinct populations of
SNe~Ia by examining the early-phase intrinsic $(B - V)_0$ color evolution of a dozen
SNe~Ia discovered very young. The "blue'' group 
exhibits blue colors that evolve slowly, while the "red'' group is characterized by red colors
and evolves more rapidly, as shown in their Figure 2 (Stritzinger et al. 2018a).

In Figure 5, we replot their Figure 2 using a subset of the sample from Stritzinger et al. (2018a)
of those SNe~Ia that were discovered or observed by LOSS, including SN~2009ig, SN~2012cg, SN~2013dy,
and SN~2013gy, as well as SN~2011fe (though not discovered by LOSS, KAIT/LOSS conducted follow-up
observations at early times). Using the same criterion as Stritzinger et al. (2018a), namely
to select SNe~Ia that have early $(B - V)$ color data within three days of the FFLT, we add six
additional LOSS SNe~Ia that were not included in Stritzinger et al. (2018a) but have either
old or new LOSS photometry published by Ganeshalingam et al. (2010)
and Stahl et al. (2019a), including SN~1999cp, SN~2002bo, SN~2002er, SN~2012Z, SN~2017cfd,
and SN~2017erp (though note that similar to SN~2011fe, the two objects SN~2002bo and SN~2017erp were
not discovered by LOSS, but KAIT/LOSS had early-time follow-up observations).
The parameters of these six new SNe~Ia are given in Table 3.

We then plot SN~2017cfd over this subset sample with a total of 11 SNe~Ia observed by LOSS,
as shown in Figure 5; open symbols represent the SNe from
Stritzinger et al. (2018a), while filled symbols are the six new SNe, and
SN~2017cfd is indicated with a filled star. Note that there is one data point for SN~2017cfd around
4.6 days that deviates from the evolution trend, caused by the bad quality of images on that
night, but the measurement is largely consistent within the error bar.
As one can see, with six more SNe added to the sample, we confirm that the 
two distinct early-time populations remain valid. SN~1999cp, SN~2012Z, and SN~2017cfd belong to
the "blue'' group, whose $(B - V)_0$ color evolves slowly at early times, similar to the other
SNe~Ia in the "blue'' group. On the other hand, SN~2002bo, SN~2002er, and SN~2017erp are consistent with
the "red'' group. It is also interesting to note that the newly added SN~2012Z is the only SN~Iax
in both our new sample and the sample from Stritzinger et al. (2018a), and it is consistent
with the "blue'' group.

Stritzinger et al. (2018a) discussed various processes that may be contributing to the early-phase 
emission and the distinct grouping; these include interaction with a nondegenerate companion,
the presence of high-velocity $^{56}$Ni, interaction with circumstellar material, and
opacity differences in the outer layers of the ejecta. They conclude that each explanation
has its own defects (see also Jiang et al. 2018), thus requiring further theoretical modeling
as well as gathering a larger sample of events. The LOSS discovery of SN~2017cfd is
the latest event that can be added to this sample (LOSS discovered or observed about half of this
$\sim 20$ SN~Ia sample with early-time color data), and it would be interesting to see if
this dichotomy persists with more data obtained in the future.

\subsection{Progenitor Constraints}\label{ss:constrainprojenitorsystem}

The very early-time observations constrain the emission from the ejecta, which can be used
to limit the radius of the progenitor star, or the companion star if the ejecta collide with
with a companion star (e.g., Kasen 2010). Such work has been applied
to several SNe~Ia, including SN~2011fe (Nugent et al. 2011; Bloom et al. (2012),
SN~2012cg (Silverman et al. 2012b), SN~2013dy (Zheng et al. 2013), SN~2013gy (Holmbo et al. 2019),
SN~2014J (Goobar et al. 2014), iPTF14atg (Cao et al. 2015),
SN~2015F (Im et al. 2015), SN~2017cbv (Hosseinzadeh et al. 2017), SN~2018oh (Li et al. 2019),
and SN~2019ein (Kawabata et al. 2019); these studies all ruled out a giant companion.


For SN~2017cfd, the earliest $B$-band observation of $\sim18.54$ mag (corrected for extinction) at 1.6\,d
limits any emission from this process to be $\nu L_{\nu}\lesssim8.7\times10^{40}$\,erg\,s$^{-1}$ at 
optical wavelengths.
Comparing these parameters with those of SN~2011fe (see Fig. 4 of Nugent et al. 2011)
and scaling the analysis to match SN~2017cfd, we obtain an upper limit
for the companion star radius to be $R_0\lesssim2.5\,{\rm R}_\sun$. This is not as stringent a constraint
as that provided by the study of SN~2011fe, which has $R_0\lesssim0.1\,{\rm R}_\sun$,
but our result for SN~2017cfd is consistent with those of other SN~Ia studies that rule out
a red-giant companion.

\section{Optical Spectra Analysis}\label{s:spectra}

We obtained a total of 12 optical spectra of SN~2017cfd
ranging from 3.5\,d to 80\,d after the FFLT.
The first spectrum was taken 13.2\,d before $B$ maximum brightness.
Figure 6 shows the full spectral sequence of SN~2017cfd.

\begin{figure*}
\centering
\includegraphics[width=.98\textwidth]{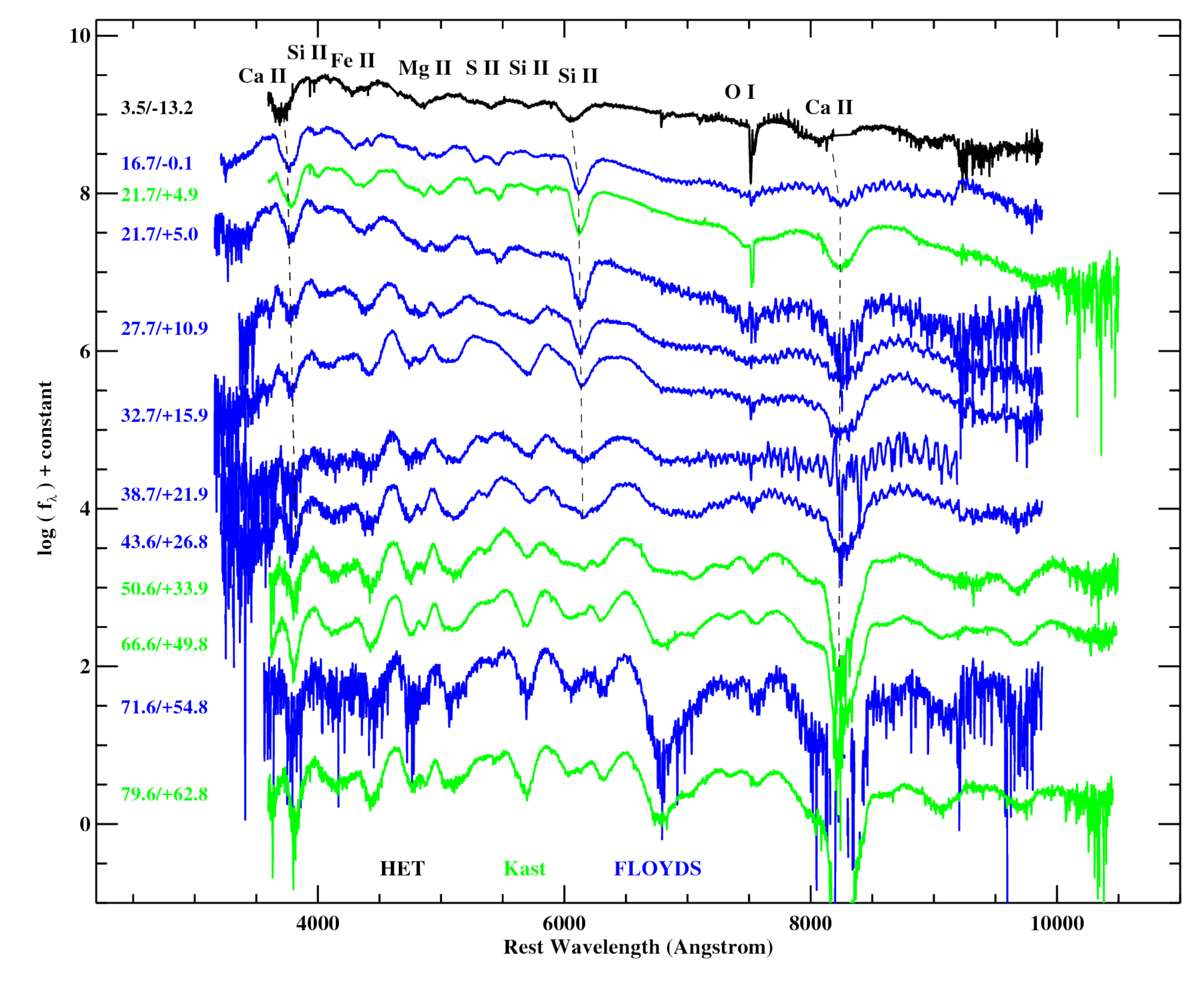}
\caption{Spectral sequence of SN~2017cfd. Each spectrum is labeled with its age relative to both the 
  FFLT and to $B$-band maximum light. Some major spectral features are labeled at the top.
  Spectra taken by different instruments are shown in different colors. In the first and third
   spectra from the top, telluric absorption is visible near 7600\,\AA\ (and a little near
   6860\,\AA\ in the first spectrum).
  Dashed lines are meant to help guide the eye when examining absorption features.
  }
\label{fig6}
\end{figure*}

We use the SuperNova IDentification code (SNID; Blondin \& Tonry 2007) to spectroscopically classify SN~2017cfd.
For nearly all of the spectra, we find that SN~2017cfd is very similar to many
normal SNe~Ia.
Thus, we conclude that SN~2017cfd is a spectroscopically normal SN~Ia, consistent with the
photometric analysis given in Section 3.

We examine the \ion{Na}{1}~D absorption feature, which is often converted into reddening (but with large scatter) through an empirical relationship (Poznanski et al. 2011; Stritzinger et al. 2018b). In several of our spectra with good signal-to-noise ratio, we clearly detect the blended \ion{Na}{1}~D at the redshifted wavelength of SN~2017cfd, but not at the rest-frame wavelength. From these spectra, we measure an averaged EW of \ion{Na}{1}~D to be ${\rm EW} = 1.72 \pm 0.18$\,\AA\ from the host galaxy. Using the best-fit relation from Stritzinger et al. (2018b), $A_V = (0.78 \pm 0.15) \times$ EW(\ion{Na}{1}~D), we estimate a host-galaxy extinction of $A_V = 1.34 \pm 0.40$\,mag, which corresponds to $E(B-V) = 0.45 \pm 0.13$\,mag assuming $R_V = 3.1$. The foreground Milky Way extinction, on the other hand, is very small according to Schlafly \& Finkbeiner (2011), only $E(B-V) = 0.02$\,mag, consistent with the nondetection of \ion{Na}{1}~D in our spectra. However, since the \ion{Na}{1}~D EW method has large scatter, and the SN would appear to be too luminous if we adopt $A_V = 1.34$\,mag estimated from the \ion{Na}{1}~D EW method, we instead adopt the extinction estimated from MLCS2k2 fitting with $A_V = 0.39 \pm 0.03$\,mag and $R_V = 1.7$ as discussed in Section 3.2. Nevertheless, the \ion{Na}{1}~D EW method serves as independent evidence showing that SN~2017cfd suffers a certain amount of host-galaxy extinction.

The spectra of SN~2017cfd exhibit absorption features from ions typically seen in SNe~Ia including
\ion{Ca}{2}, \ion{Si}{2}, \ion{Fe}{2}, \ion{Mg}{2}, \ion{S}{2}, and \ion{O}{1}.
We do not find a clear \ion{C}{2} feature, which is
seen in over one-fourth of all SNe~Ia (e.g., Parrent et al. 2011; Thomas et al. 2011;
Folatelli et al. 2012; Silverman \& Filippenko 2012),
in the earliest spectrum of SN~2017cfd at phase $-13.2$\,d. Although
there appears to be a suspicious dip at the red edge of \ion{Si}{2} $\lambda$6355 that might be caused
by \ion{C}{2} $\lambda$6580, this feature is very weak, not as clear as those seen in a few
other SN~Ia early-time spectra such as SN~2013dy (Zheng et al. 2013) and SN~2017cbv (Hosseinzadeh et al. 2017).
Strong absorption features of \ion{Si}{2}, including \ion{Si}{2} $\lambda$4000,
\ion{Si}{2} $\lambda$5972, and \ion{Si}{2} $\lambda$6355, are clearly seen in all spectra,
though the \ion{Si}{2} $\lambda$5972 feature in SN~2017cfd is relatively weak.

\begin{figure}
\centering
\includegraphics[width=.48\textwidth]{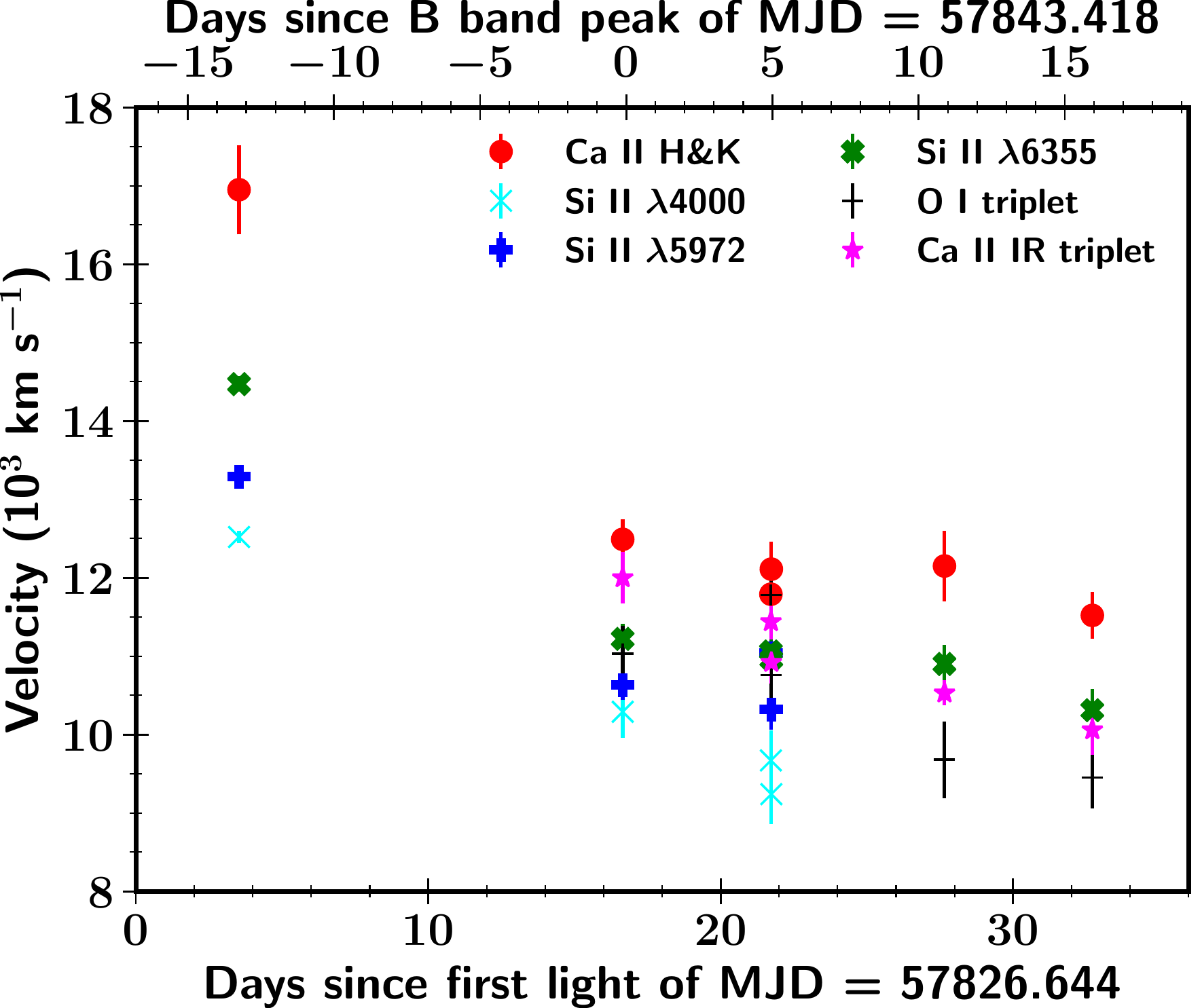}
\caption{The expansion velocity evolution of different lines measured
  from the spectra of SN~2017cfd.
  }
\label{fig7}
\end{figure}

We then measure the individual line velocities from the minimum of the absorption features
(see Silverman et al. 2012c, for details) and show them in Figure 7.
As expected, the velocities of all lines decrease from early phases to
relative constancy around peak brightness, as seen in almost all SNe~Ia.
Specifically, the velocity of the \ion{Si}{2}$\lambda$6355 line decreases from $\sim14,500$\,\kms\ 
at discovery to $\sim11,200$\,\kms\ around maximum light, and then continues to
decrease thereafter. 
\ion{Si}{2}$\lambda$6355 also has the largest velocity among all three
\ion{Si}{2} lines ($\lambda$4000, $\lambda$5792, and $\lambda$6355).
But \ion{Ca}{2} {H\&K} tends to exhibit the highest velocity among all the features, higher
even than the \ion{Ca}{2} near-infrared triplet, consistent with most SNe~Ia.
The strong absorption of \ion{Si}{2}$\lambda$6355 is commonly used to estimate the photospheric velocity;
$\sim11,200$\,\kms\ at peak brightness is very typical of normal SNe~Ia
(e.g., Wang et al. 2013; Stahl et al. 2019b, submitted.).


\section{Conclusions}\label{s:conclusions}

In this paper we have presented optical photometric and spectroscopic observations
of SN~2017cfd, which was discovered very young,
with the first detection merely $1.6 \pm 0.7$\,d after the FFLT.
We find that SN~2017cfd is a normal SN~Ia in nearly every respect.
(1) SN~2017cfd took $\sim16.8$\,d to reach $B$-band maximum, typical of SNe~Ia.
(2) There is a certain amount of host-galaxy extinction of SN~2017cfd
based on the detection of \ion{Na}{1}~D lines from the host galaxy as well as the MLCS2k2
light-curve fitting method; however, considering that the \ion{Na}{1}~D EW method has 
large scatter, we adopt the extinction estimated from MLCS2k2 fitting
with $A_V = 0.39 \pm 0.03$\,mag and $R_V = 1.7$. After extinction correction,
its maximum brightness has a normal luminosity, $B = -19.2 \pm 0.2$\,mag.
(3) An estimated $\Delta m_{15}(B)$ value of 1.16\,mag
along with spectral information supports its normal SN~Ia classification.
(4) SN~2017cfd has a \ion{Si}{2} $\lambda$6355 velocity of $\sim11,200$\,\kms\ at peak 
brightness, also very typical of normal SNe~Ia.

SN~2017cfd was detected very early. There are currently fewer
than $\sim 20$ SNe~Ia with color data in the first three days, and fewer
than a dozen SNe~Ia with color data in the first two days.
Using the early-time photometry, we are able to constrain the companion star
radius to be $\lesssim2.5\,{\rm R}_\sun$, ruling out a red-giant companion associated with
SN~2017cfd.
We also find that the intrinsic $(B - V)_0$ color evolution of SN~2017cfd at very 
early times belongs to the "blue'' population,
consistent with the dichotomy of the "red'' and "blue'' populations at early phases.
Therefore, thanks to the early discovery and photometric follow-up, SN~2017cfd remains
valuable at this stage for building up a bigger sample
for studying SNe~Ia at very early times. A significantly larger sample is being obtained with
available new facilities (e.g., the Zwicky Transient Facility; ZTF), and with future new
telescopes there will be many such discoveries (e.g., the Large Synoptic Survey Telescope). 

During the final completion stages of this manuscript, Yao et al. (2019)
released high-quality light curves of 127 SNe~Ia discovered by ZTF in 2018, 
and a large fraction of their sample may have $(g - r)$ color data in the first 
three days. However, since their analysis use phase measurements relative to the 
$g$-band maximum instead of the FFLT, a direct comparison comparison cannot be
made with our analysis. A reanalysis to find the FFLT from the Yao et al. (2019) 
sample is required but is beyond the scope of this paper.

\begin{acknowledgments}

A.V.F.'s group at U.C. Berkeley is grateful for financial assistance
from Gary \& Cynthia Bengier, the Christopher R. Redlich Fund, the TABASGO Foundation,
and the Miller Instute for Basic Research in Science (U.C. Berkeley).
We thank the staffs of the various observatories at which data were obtained.
This work makes use of observations from the LCO network.
KAIT and its ongoing operation were made possible by donations from
Sun Microsystems, Inc., the Hewlett-Packard Company, AutoScope
Corporation, Lick Observatory, the NSF, the University of California, the Sylvia
\& Jim Katzman Foundation, and the TABASGO Foundation.
Research at Lick Observatory is partially supported by a generous gift from Google.
This work is part of the project "Transient Astrophysical Objects'' GINOP 2.3.2-15-2016-00033 of the 
National Research, Development, and Innovation Office (NKFIH), Hungary, funded by the European Union.

\end{acknowledgments}


\begin{thebibliography}{50}
\expandafter\ifx\csname natexlab\endcsname\relax\def\natexlab#1{#1}\fi

%
%
\bibitem[{Benetti}{et~al.}(2004)]{Benetti04} 
Benetti, S., Meikle, P., Stehle, M., et al. 2004, MNRAS, 348, 261

%
%
%
\bibitem[{Blondin}{Tonry}(2007)]{Blondin07}
Blondin, S., \& Tonry, J.~L. 2007, \apj, 666, 1024

\bibitem[{Bloom}{et~al.}(2012)]{Bloom12}  Bloom, J., Kasen, D., Shen, K., et al. 2012, ApJL, 744, L17

%
\bibitem[{Brown}{et~al.}(2013)]{Brown13} Brown, T. M., Baliber, N., Bianco, F. B., et al. 2013, PASP, 125, 1031

\bibitem[{Brown}{et~al.}(2019)]{Brown19} Brown, P. J., Hosseinzadeh, G., Jha, S. W., et al. 2019, ApJ, 877, 152

\bibitem[{Cao}{et~al.}(2016)]{Cao16} Cao, Y., Kulkarni, S. R., Gal-Yam, A., et al. 2016, ApJ, 832, 86

\bibitem[{Cao}{et~al.}(2015)]{Cao15} Cao, Y., Kulkarni, S. R., Howell, D. A., et al. 2015, Nature, 521, 328

%
%
%
%
%
\bibitem[{Chonis}{et~al.}(2014)]{Chonis14} Chonis, T. S., Hill, G. J., Lee, H., et al. 2014, SPIE, 91470A, 1

\bibitem[{Chonis}{et~al.}(2016)]{Chonis16}  Chonis, T. S., Hill, G. J., Lee, H., et al. 2016, SPIE, 99084C, 1

%
%
%
%
\bibitem[{de Vaucouleurs}{et~al.}(1991)]{deVaucouleurs91}  
de Vaucouleurs, G., de Vaucouleurs, A., Corwin, H. G., Jr.,
Buta, R. J., Paturel, G., \& Fouqu\'{e}, P. 1991, RC3.9, "Third Reference 
Catalogue of Bright Galaxies,'' version 3.9 (New York: Springer)

%
\bibitem[{Falco}{et~al.}(1999)]{Falco99} Falco, E. E., Kurtz, M. J., Geller, M. J., et al. 1999, PASP, 111, 438

\bibitem[{Filippenko}(1982)]{Filippenko82} Filippenko, A. V. 1982, PASP, 94, 715

\bibitem[{Filippenko}(1997)]{Filippenko97} Filippenko, A. V. 1997, ARA\&A, 35, 309

\bibitem[{Filippenko}(2005)]{Filippenko05} Filippenko, A. V. 2005, in 1604--2004, Supernovae
    as Cosmological Lighthouses, ed. M. Turatto, et al.
    (San Francisco: ASP), 87

\bibitem[{Filippenko} {et~al.}(2001)]{Filippenko01}
 Filippenko, A.~V., Li, W.~D., Treffers, R.~R., \& Modjaz, M. 2001, in
 Small-Telescope Astronomy on Global Scales., ed. B.~Paczy\'{n}ski, W.~P.
 Chen, \& C.~Lemme (San Francisco: ASP), 121

%
%
%
%
%
%
%
\bibitem[{Foley} {et~al.}(2018)]{Foley18} Foley, R. J., Scolnic, D., Rest, A., et al. 2018, MNRAS, 475, 193

%
\bibitem[{Ganeshalingam} {et~al.}(2010)]{Ganeshalingam10} Ganeshalingam, M., Li, W., Filippenko, A. V., et~al. 2010, ApJS, 190, 418

\bibitem[{Goobar} {et~al.}(2014)]{Goobar14} Goobar, A., Johansson, J., Amanullah, R., et al. 2014, 784, L12

%
\bibitem[{Guevel} {et~al.}(2017)]{Guevel17} Guevel, D., \& Hosseinzadeh, G. 2017, dguevel/PyZOGY: Initial release (Version v0.0.1). Zenodo.

\bibitem[{Guy} {et~al.}(2007)]{Guy07} Guy, J., Astier, P., Baumont, S., et al. 2007, A\&A, 466, 11

%
\bibitem[{Hachisu} {et~al.}(1996)]{Hachisu96}  Hachisu, I., Kato, M., \& Nomoto, K. 1996, ApJL, 470, L97

%
\bibitem[{Hicken} {et~al.}(2009)]{Hicken09} Hicken, M., Wood-Vasey, W. M., Blondin, S., et al. 2009, ApJ, 700, 1097

\bibitem[{Hillebrandt} {Niemeyer}(2000)]{Hillebrandt00}  Hillebrandt, W., \& Niemeyer, J. C. 2000, ARA\&A, 38, 191

%
\bibitem[{Holmbo} {et~al.}(2019)]{Holmbo19}  Holmbo, S., Stritzinger, M. D., Shappee, B. J., et al. 2019, A\&A, 627, A174

\bibitem[{Hosseinzadeh} {et~al.}(2017)]{Hosseinzadeh17} Hosseinzadeh, G., Sand, D., Valenti, S., et al. 2017, 845, 11

%
\bibitem[{Hoyle} {et~al.}(1960)]{Hoyle60} Hoyle, F., \& Fowler, W. A. 1960, ApJ, 132, 565

%
%
%
\bibitem[{Iben} {Tutukov}(1984)]{Iben84} Iben, I., Jr., \& Tutukov, A. V. 1984, ApJS, 54, 335

\bibitem[{Im} {et~al.}(2015)]{Im15} Im, M., Choi, C., Yoon, S., et al. 2015, ApJS, 221, 22

\bibitem[{Jiang} {et~al.}(2018)]{Jiang18} Jiang, J., Doi, M., Maeda, K., \& Shigeyama, T., 2018, ApJ, 865, 149

\bibitem[{Jha} {et~al.}(2019)]{Jha19}  Jha, S., Maguire, K., \& Sullivan, M. 2019, Nature, 3, 706

\bibitem[{Jha} {et~al.}(2007)]{Jha07} Jha, S., Riess, A. G., \& Kirshner, R. P. 2007, ApJ, 659, 122

%
\bibitem[{Kawabata} {et~al.}(2019)]{Kawabata19} Kawabata, M., Maeda, K., Yamanaka, M., et al. 2019, arXiv:1908.03001

\bibitem[{Kasen}(2010)]{Kasen10} Kasen, D. 2010, ApJ, 708, 1025

%
%
%
%
%
%
%
%
\bibitem[{Leaman}{et~al.}(2011)]{Leaman11} Leaman, J., Li, W., Chornock, R., \& Filippenko, A. V. 2011, MNRAS, 412, 1419

%
%
\bibitem[{Li}{et~al.}(2003)]{Li03}  Li, W., Filippenko, A. V., Chornock, R., \& Jha, S. 2003, PASP, 115, 844

\bibitem[{Li}{et~al.}(2019)]{Li19}  Li, W., Wang, X., Vinko, J., et al. 2019, ApJ, 870, 12

\bibitem[{Maeda}{et~al.}(2018)]{Maeda18}  Maeda, K., Jiang, J., Shigeyama, T. \& Doi, M. 2018, ApJ, 861, 78

\bibitem[{Magee}{et~al.}(2018)]{Magee18} Magee, M. R., Sim, S. A., Kotak, R. \& Kerzendorf, W. E. 2018, A\&A, 614, A115

%
%
%
%
%
%
%
%
\bibitem[{Meng}{et~al.}(2009)]{Meng09} Meng, X., Chen, X., \& Han, Z., et al. 2009, MNRAS, 395, 2103

%
%
\bibitem[{Miller}{Stone}(1993)]{Miller93}
Miller, J.~S., \& Stone, R.~P.~S. 1993, Lick Obs. Tech. Rep. 66 (Santa
  Cruz: Lick Obs.)

%
\bibitem[{Nugent}{et~al.}(2011)]{Nugent11}  Nugent, P.~E., Sullivan, M., Cenko, S. B., et al. 2011, Nature, 480, 344

%
\bibitem[{Pakmor}{et~al.}(2012)]{Pakmor12} Pakmor, R., Kromer, M., Taubenberger, S., Sim, S. A., R\"{o}pke, F. K., \& Hillebrandt, W. 2012, ApJL, 747, L10

\bibitem[{Papadogiannakis}{et~al.}(2019)]{Papadogiannakis19} Papadogiannakis, S., Goobar, A., Amanullah, R., et al. 2019, MNRAS, 483, 5045

%
%
\bibitem[{Perlmutter}{et~al.}(1999)]{Perlmutter99} Perlmutter, S., Aldering, G., Goldhaber, G., et~al. 1999, ApJ, 517, 565

\bibitem[{Phillips}{et~al.}(1993)]{Phillips93} Phillips, M. M. 1993, ApJ, 413, L105

%
\bibitem[{Pignata}{et~al.}(2004)]{Pignata04} Pignata, G., Patat, F., Benetti, S., et al. 2004, MNRAS, 355, 178

%
\bibitem[{Piro}{et~al.}(2013)]{Piro13} Piro, A. L., \& Nakar, E. 2013, ApJ, 769, 67

\bibitem[{Piro}{et~al.}(2014)]{Piro14} Piro, A. L., \& Nakar, E. 2014, ApJ, 784, 85

\bibitem[{Polin}{et~al.}(2019)]{Polin19} Polin, A., Nugent, P. \& Kasen, D. 2019, ApJ, 873, 84

\bibitem[{Poznanski} {et~al.}(2011)]{Poznanski11}
Poznanski, D., Ganeshalingam, M., Silverman, J.~M., \& Filippenko,
  A.~V. 2011, MNRAS, 415, L81

\bibitem[{Rabinak} {et~al.}(2012)]{Rabinak12} Rabinak, I., Livne, E., \& Waxman, E. 2012, ApJ, 757

\bibitem[{Riess} {et~al.}(2018)]{Riess18} Riess, A.~G., Casertano, S., Yuan, W., et~al. 2018, ApJ, 855, 136

\bibitem[{Riess} {et~al.}(2019)]{Riess19} Riess, A.~G., Casertano, S., Yuan, W., et~al. 2019, ApJ, 876, 85

\bibitem[{Riess} {et~al.}(1998)]{Riess98} Riess, A.~G., Filippenko, A. V., Challis, P., et~al. 1998, AJ, 116, 1009

%
%
\bibitem[{Ropke} {et~al.}(2012)]{Ropke12} R\"{o}pke, F. K., Kromer, M., Seitenzahl, I. R., et al. 2012, ApJL, 750, L19

%
\bibitem[{Scalzo} {et~al.}(2019)]{Scalzo19} Scalzo, R. A., Parent, E., Childress, M., et al. 2019, MNRAS, 483, 628

\bibitem[{Schlafly} {Finkbeiner}(2011)]{Schlafly11} Schlafly, E. F., \& Finkbeiner, D. P. 2011, ApJ, 737, 103

%
%
%
\bibitem[{Shivvers} {et~al.}(2019)]{Shivvers19} Shivvers, I., Filippenko, A. V., Silverman, J.~M., et al. 2019, MNRAS, 482, 1545

%
%
\bibitem[{Silverman} {et~al.}(2012)]{Silverman12a} Silverman, J. M., Foley, R., Filippenko, A. V., et al. 2012a, MNRAS, 425, 1789

\bibitem[{Silverman} {et~al.}(2012)]{Silverman12b} Silverman, J.~M., Ganeshalingam, M., Cenko, S. B., et~al. 2012b, ApJ, 756, L7

\bibitem[{Silverman} {et~al.}(2012)]{Silverman12c} Silverman, J.~M., Kong, J. J., \& Filippenko, A. V. 2012c, MNRAS, 425, 1819

%
%
\bibitem[{Stahl} {et~al.}(2019)]{Stahl19a}  Stahl, B. E., Zheng, W., de Jaeger, T., et al. 2019a, in press (arXiv:1909.11140)

\bibitem[{Stahl} {et~al.}(2019)]{Stahl19b} Stahl, B. E., Zheng, W., de Jaeger, T., et al. 2019b, submitted

%
%
%
\bibitem[{Stetson} {et~al.}(1987)]{Stetson87} Stetson, P. B. 1987, PASP, 99, 191

\bibitem[{Stritzinger} {et~al.}(2018)]{Stritzinger18a} Stritzinger, M. D., Shappee, B., Piro, A., et al. 2018a, ApJ, 864, L35

\bibitem[{Stritzinger} {et~al.}(2018)]{Stritzinger18b} Stritzinger, M. D., Taddia, F., Burns, C. R., et al. 2018b, A\&A, 609, 135

\bibitem[{Stritzinger} {et~al.}(2015)]{Stritzinger15} Stritzinger, M. D., Valenti, S., Hoeflich, P., et al. 2015, A\&A, 573, 2

%
%
%
\bibitem[{Tonry} {et~al.}(2012)]{Tonry12} Tonry, J. L., Stubbs, C. W., Lykke, K. R., et al. 2012, ApJ, 750, 99

%
%
%
%
\bibitem[{Valenti} {et~al.}(2016)]{Valenti16} Valenti, S., Howell, D. A., Stritzinger, M. D., et al. 2016, MNRAS, 459, 3939

%
%
%
%
%
%
%
\bibitem[{Wang}{et~al.}(2013)]{Wang13} Wang, X., Wang, L., Filippenko, A. V., Zhang, T., \& Zhao, X. 2013, Science, 340, 170

\bibitem[{Webbink}{et~al.}(1984)]{Webbink84} Webbink, R. F. 1984, ApJ, 277, 355

%
%

\bibitem[{Yao}{et~al.}(2019)]{Yao19} Yao, Y., Miller, A. A., Kulkarni, S. R., et al. 2019, arXiv:1910.02967

\bibitem[{Zackay}{et~al.}(2016)]{Zackay16}Zackay, B., Ofek, E. O., \& Gal-Yam, A. 2016, ApJ, 830, 27

%
%
\bibitem[{Zheng}{et~al.}(2017a)]{zheng17a}Zheng, W., \& Filippenko A. V. 2017, ApJ, 838, L4

\bibitem[{Zheng}{et~al.}(2017b)]{zheng17b}Zheng, W., Filippenko A. V., Mauerhan, J., et al. 2017a, ApJ, 841, 64

\bibitem[{Zheng}{et~al.}(2017c)]{zheng17c}Zheng, W., Kelly, P., \& Filippenko, A. V. 2017b, ApJ, 848, 66

\bibitem[{Zheng}{et~al.}(2018)]{zheng18}Zheng, W., Kelly, P., \& Filippenko, A. V. 2018, ApJ, 858, 104

\bibitem[{Zheng}{et~al.}(2014)]{zheng14} Zheng, W., Shivvers, I., Filippenko, A. V., et al. 2014, ApJ, 783, L24

\bibitem[{Zheng}{et~al.}(2013)]{zheng13} Zheng, W., Silverman, J. M., Filippenko, A. V., et al. 2013, ApJ, 778, L15

%

\end{thebibliography}
\end{document}